\documentclass[10pt, conference, compsocconf, letterpaper]{IEEEtran}

\usepackage[numbers]{natbib}

\usepackage{hyperref}
\usepackage{url}
\usepackage{placeins} 

\usepackage{boxedminipage}
\usepackage{listings}
\usepackage{subfigure}

\usepackage{amsmath}
\usepackage{amsfonts}
\usepackage{enumerate}
\usepackage{graphicx}

\usepackage{relsize}
\usepackage{xspace}

\newcommand{\mparam}{\mathbb{M_P}}
\newcommand{\mwithoutparam}{\mathbb{M_W}}

\newcommand{\visitorclass}[1]{\ensuremath{vis(\mbox{#1})}}
\newcommand{\auxsymbol}{\mathit{aux}}
\newcommand{\auxname}[1]{\ensuremath{\auxsymbol(}#1\ensuremath{)}}

\date{}

\newcommand{\tab}{\phantom{--}}

\newcommand{\xcode}[1]{{\relsize{-1}\textsf{#1}}}

\begin{document}

\title{Transformations between  Composite\\ and Visitor implementations in Java}

\author{\IEEEauthorblockN{Akram Ajouli}
\IEEEauthorblockA{ASCOLA (EMN - INRIA - LINA)\\
Nantes, France\\
Akram.Ajouli@mines-nantes.fr}
\and 

\IEEEauthorblockN{Julien Cohen}
\IEEEauthorblockA{LINA (UMR 6241, CNRS, Univ. Nantes, EMN)\\
Nantes, France\\
Julien.Cohen@univ-nantes.fr}
\and 

\IEEEauthorblockN{Jean-Claude Royer}
\IEEEauthorblockA{ASCOLA (EMN - INRIA - LINA)\\
Nantes, France\\
Jean-Claude.Royer@mines-nantes.fr}}

%
\setcounter{page}{1}
\pagenumbering{arabic}

\maketitle


\newcommand{\intellij}{IntelliJ \textsc{Idea}\xspace}

\begin{abstract}

Basic automated refactoring operations can be chained to
perform complex structure transformations.
This is useful for recovering the initial architecture of a
source code which has been degenerated with successive
evolutions during its maintenance lifetime.
This is also useful for changing the structure of a program
so that a maintenance task at hand becomes modular when it
would be initially  crosscutting.

We focus on programs structured according to
Composite and Visitor design patterns, which have dual
properties with respect to modularity.
We consider a refactoring-based round-trip transformation between
these two structures and we study how that transformation
is impacted by four variations in the implementation of these patterns.
We validate that study by computing the smallest
preconditions for the resulting transformations.
We also automate the transformation and apply it to
JHotDraw, where the studied variations occur.

\end{abstract}


\IEEEpeerreviewmaketitle

\section{Introduction}
\label{sect:introduction}


The complexity of program maintenance depends on the quality
of its structure. 
But, for a maintenance task at hand, some design choices 
that cannot be qualified as good or bad could also impact that
complexity.
 This is illustrated by the case of Composite and Visitor
patterns. Despite their usefulness in facilitating reuse and
maintainability, each one is more suitable for a specific kind of
maintenance. While the Composite (as well as Interpreter
pattern and simple class hierarchies) offers modular
maintenance with respect to data types, the Visitor pattern provides
modular maintenance with respect to functions~\cite{Gamma:1995}.
These two patterns can be good at design time because you usually do not  know at
that moment if you will face more maintenance on the data
axis or on the function axis in the future.
These (micro-)architectures are complementary.
%
%
Automatic switching between the two structures at the source
code level allows to benefit from the best
pattern with respect to a maintenance task at
hand~\cite{Cohen-Douence-Ajouli:2012}. 


Such a behavior-preserving transformation is given
by Ajouli~\cite{Akram-Ciel-2012} and has been validated by a static
analysis by Cohen and Ajouli~\cite{Cohen-Ajouli-SAC}.
However, that transformation is designed for a given
implementation of the pattern (methods with no
parameters, no returned values, abstract class at the root
of the composite structure and a single level in the composite
hierarchy).
Since various design choices for implementing those patterns are
found in real softwares, the described transformation
does not apply directly.


%

Our contribution in this paper is the extension of the transformation presented by Ajouli~\cite{Akram-Ciel-2012} to take into account 
four variations in the 
implementations of the Composite and Visitor patterns. We validate the built transformations by computing their minimum preconditions and 
by applying them to JHotDraw in which the four variations occur.  
%

For each variation considered on the Composite pattern, we discuss how it is reflected on the
dual Visitor implementation, on the round-trip
transformation and on its preconditions (Sec.~\ref{sect:variation}). 
Then, we validate these changes in the transformation
algorithms by applying them to JHotDraw both practically and formally (Sec.~\ref{usecase}). 
Before addressing the variations of the patterns and transformations, we review the basic transformation on a toy example in Sec.~\ref{sect:basictrasnformation}.

%
%

\section{Transformation between Composite and Visitor}
\label{sect:basictrasnformation}


To illustrate each pattern variation, we start with a toy
implementation of the Composite pattern, given in Fig.~\ref{base-prog:data}.
This program is composed by an abstract
class, \xcode{Figure}, with two declared business
methods, \xcode{print} and \xcode{show}, and two
subclasses.  One of these classes, \xcode{Group}, contains
references to \xcode{Figure} objects, and the business
methods in that classes make recursive invocations on those
objects.
This is a simple implementation of the Composite pattern. Fig.~\ref{base-prog:state3-visitor} gives a program with the
same semantics but which implements the Visitor pattern.

\begin{figure}[!htp]
\begin{minipage}{\columnwidth}

\begin{lstlisting}
abstract class Figure {

  abstract  void print();

  abstract  void show();}
\end{lstlisting}
\begin{lstlisting}
class Rectangle extends Figure{

 void print(){System.out.println("Rectangle");}

 void show(){System.out.println("Rectangle:" + this);}}
\end{lstlisting}
\end{minipage}

\begin{minipage}{\columnwidth}
\begin{lstlisting}[firstline=3]
class Group extends Figure {
 ArrayList<Figure> children 
        = new ArrayList<Figure> () ;
 void print() {
   System.out.println("Group:");
   for (Figure child : children) {child.print();}}

 void show(){
   System.out.println("Group " + this);
   for (Figure child : children) {child.show();}
   System.out.println("(end)");}}
\end{lstlisting}
\end{minipage}

\caption{A program structured according to Composite Design pattern.}
\label{base-prog:data}
\end{figure}

\begin{figure}[!h]
\subfigure[Data classes.]{
\begin{minipage}{\columnwidth}
\begin{lstlisting}
abstract class Figure {
 void print(){accept(new PrintVisitor());}

 void show(){accept(new ShowVisitor());}

 abstract void accept(Visitor v);}
\end{lstlisting}
\begin{lstlisting}
class Rectangle extends Figure{

 void accept(Visitor v){v.visit(this);}}
\end{lstlisting}
\begin{lstlisting}[firstline=3]
class Group extends Figure {
 ArrayList<Figure> children = new ArrayList<Figure>();

 void accept(Visitor v) {v.visit(this);}}
\end{lstlisting}

\end{minipage}

\label{composites}
}
\hfill
\subfigure[Visitor classes.]{
\begin{minipage}{\columnwidth}
\begin{lstlisting}
abstract class Visitor {

 abstract void visit(Rectangle r);

 abstract void visit(Group g);}
\end{lstlisting}
\begin{lstlisting}
class PrintVisitor extends Visitor {
 void visit(Rectangle r){System.out.println("Rectangle");}

 void visit(Group g) {
  System.out.println("Group:");
  for (Figure child : g.children){child.accept(this);}}}
\end{lstlisting}
\begin{lstlisting}
class ShowVisitor extends Visitor{
 void visit(Rectangle r){
                System.out.println("Rectangle:"+r);}
 void visit(Group g){
  System.out.println("Group "+g);
  for (Figure child : g.children){child.accept(this);}	
  System.out.println("(end)");}}
\end{lstlisting}

\end{minipage} 
\label{visitors}
}

\caption{Visitor structure of the program shown in the Fig.~\ref{base-prog:data}.}
\label{base-prog:state3-visitor}
\end{figure}

We consider the two transformations given in
Fig.~\ref{transfo-algorithms}
from~\cite{Akram-Ciel-2012} to
switch between these two implementations of the Composite
and Visitor structures.
These transformations are built by composing elementary
behavior-preserving refactoring operations from \intellij (a similar transformation is also possible with the refactoring operations of Eclipse).
This composition is based on a meaningful orchestration of
these operations in order to get the right structure (briefly explain below).
The chain of operations is automated by using the API of \intellij (some refactoring
operations are extended or modified in order to satisfy the
fully automation of the
transformation). 

%
%


%
%

We use the following notations to abstract the algorithm from the given example:

\begin{itemize}
\item $\mathbb{M}$: set of business methods, here $\mathbb{M} = $\{\xcode{print},\xcode{show}\}.
\item $\mathbb{C}$: set of Composite hierarchy classes except its root, here $\mathbb{C} = $\{\xcode{Rectangle}, \xcode{Group}\} 


\item $S$: root of the Composite hierarchy, here $S =$ \xcode{Figure}.
\item $vis$: function that generates a visitor class name from a business method name, here \visitorclass{\xcode{print}} $=$ \xcode{PrintVisitor}. 
\item $\mathbb{V}$: set of visitor classes, here $\mathbb{V}=\{ \visitorclass{$m$}\}_{m \in \mathbb{M} } = $\{\xcode{PrintVisitor}, \xcode{ShowVisitor}\}. 
\item  $\auxsymbol$: function used to generate names of temporary methods, here $\auxname{$\xcode{print}$}=$ \xcode{printAux}. 

\end{itemize}



%

\newcommand{\operation}[1]{\textbf{#1}}

\begin{figure}

\subfigure[Base Composite$\rightarrow$Visitor transformation.]{


\begin{boxedminipage}{\columnwidth}

\begin{enumerate}[1)]

\relsize{-1}
\sf

\item \label{algo-composite-visitor:create-visitors} 
ForAll m in $\mathbb{M}$ do   
  \operation{CreateEmptyClass}(\visitorclass{m})\\[-2mm]

\item   \label{algo-composite-visitor:create-indirection} 
ForAll m in $\mathbb{M}$ do 
   \operation{CreateIndirectionInSuperClass}(S,m, \auxname{m})\\[-2mm]

\item  \label{algo-composite-visitor:inline-methods} 
 ForAll m in $\mathbb{M}$, c in $\mathbb{C}$ do \operation{InlineMethodInvocations}(c, m, \auxname{m}) \\[-2mm]

\item  \label{algo-composite-visitor:add-parameter} 
 ForAll m in $\mathbb{M}$ do \operation{AddParameterWithReuse}(S, \auxname{m}, \visitorclass{m})\\[-2mm]

\item  \label{algo-composite-visitor:move} 
 ForAll m in $\mathbb{M}$, c in $\mathbb{C}$ do \operation{MoveMethodWithDelegate}(c, \auxname{m}, \visitorclass{m}, "visit")\\[-2mm]

\item  \label{algo-composite-visitor:superclass} 
 \operation{ExtractSuperClass}($\mathbb{V}$, "Visitor")\\[-2mm]

\item  \label{algo-composite-visitor:generalise-parameter} 
 ForAll m in $\mathbb{M}$ do \operation{UseSuperType}(S, \auxname{m}, \visitorclass{m}, "Visitor")\\[-2mm]

\item  \label{algo-composite-visitor:merge} 
  \operation{MergeDuplicateMethods}(S, \{\auxname{m} \}$_{\mbox{m} \in \mbox{$\mathbb{M}$}}$, "accept")

\end{enumerate}

\end{boxedminipage}

\label{fig-composite-visitor-transfo}
}
\hfill
\subfigure[Base Visitor$\rightarrow$Composite transformation.]{
\begin{boxedminipage}{\columnwidth}
\begin{enumerate}[I )]

\relsize{-1}

\sf


\item \label{visitor-composite-algo:specialize-accept}
   ForAll v in $\mathbb{V}$ do \operation{AddSpecializedMethodInHierarchy}(S,\\ \tab \tab"accept","Visitor",v)\\[-2mm]

\item \label{visitor-composite-algo:delete-method-in-hierarchy}  \operation{DeleteMethodInHierarchy}(S,accept,"Visitor")\\[-2mm]

\item \label{visitor-composite-algo:pushdown-visit}
       ForAll c in $\mathbb{C}$ do \operation{PushDownAll}("Visitor","visit",c)\\[-2mm]
  
\item \label{visitor-composite-algo:inline-visit}
       ForAll v in $\mathbb{V}$, c in $\mathbb{C}$ do \operation{InlineMethod}(v,"visit",c)        \\[-2mm]

\item \label{visitor-composite-algo:rename-aux}
       ForAll m in $\mathbb{M}$ do \operation{RenameMethod}(S,accept,\visitorclass{m},\auxname{m})\\[-2mm]

\item \label{visitor-composite-algo:remove-param}
       ForAll m in $\mathbb{M}$ do \operation{RemoveParameter}(S,\auxname{m},\visitorclass{m})\\[-2mm]

\item \label{visitor-composite-algo:fold}
       ForAll m in $\mathbb{M}$ do \operation{ReplaceMethodDuplication}(S,m)\\[-2mm]
      
\item \label{visitor-composite-algo:pushdown-m}
       ForAll m in $\mathbb{M}$ do \operation{PushDownImplementation}(S,m)\\[-2mm]

\item \label{visitor-composite-algo:pushdown-aux}
       ForAll m in $\mathbb{M}$ do \operation{PushDownAll}(S,\auxname{m})\\[-2mm]

\item \label{visitor-composite-algo:inline-aux}
       ForAll m in $\mathbb{M}$, c in $\mathbb{C}$ do \operation{InlineMethod}(c,\auxname{m})\\[-2mm]
 
\item \label{visitor-composite-algo:delete-visitors}
       ForAll v in $\mathbb{V}$ do \operation{DeleteClass}(v)\\[-2mm]
    
\item \label{visitor-composite-algo:delete-visitor}
      \operation{DeleteClass}("Visitor")

\end{enumerate}

\end{boxedminipage}
\label{fig-algo-retour}
}

\caption{Base Algorithms for reversible transformation from Composite to Visitor.}
\label{transfo-algorithms}

\end{figure}

\subsection{Composite to Visitor}
\label{FCompToVisitor}

         The Composite$\rightarrow$Visitor algorithm of
         Fig.~\ref{fig-composite-visitor-transfo} is
         explained with three stages: preparing for moving
         business code (steps 1 to 4); moving the business
         code to the Visitor classes (step 5) and
         recovering the conventional structure of the
         Visitor pattern (steps 6 to 8).

          \paragraph{Steps 1 to 4:  Preparing for moving business code} 
First, we create an empty visitor class
(step~\ref{algo-composite-visitor:create-visitors}) for each
business method of the program (\xcode{PrintVisitor} and \xcode{ShowVisitor}).
Then, in order to preserve the interface, we introduce a
delegator for the business methods
(step~\ref{algo-composite-visitor:create-indirection}). The
initial business method is now split into
the \emph{delegator} which keeps the name of the business
methods, its type and its defining class and
the \emph{business code} contained in the deleguee
method. In the following, we call \emph{auxiliary} methods those
deleguee methods that contain the business code.

The introduction of delegators have transformed direct
recursive invocations into indirect recursion. We replace
invocations of delegators by invocations of auxiliary
methods in business code to recover direct recursion and
avoid using the delegator in business code
(step~\ref{algo-composite-visitor:inline-methods}).

%
Finally, to be able to move business code to the right visitor
classes, we introduce into each auxiliary method its
target class by adding to it the visitor class type and name as a dummy parameter
(step~\ref{algo-composite-visitor:add-parameter}).

         \paragraph{Step 5: Moving business code to visitor classes} 

Move auxiliary methods containing the business code to
visitor classes and rename them into "visit". The famous
double dispatch involved in the Visitor pattern is created
by keeping (again) a delegator in the originating class for
each moved method.

         \paragraph{Steps 6 to 8: Recovering Visitor structure}

We extract a super-class for visitor classes
(step~\ref{algo-composite-visitor:superclass}). 
That class
will contain the abstract declarations of the \emph{visit}
methods.
In the delegators (which are in the Composite
side), we change the type of the parameter to the new super-class \xcode{Visitor}
(step~\ref{algo-composite-visitor:generalise-parameter}).
Finally, we unify all delegators in Composite side into a single method 
\emph{accept}
 (step~\ref{algo-composite-visitor:merge}).\footnote{\operation{MergeDuplicateMethods}
  and \operation{AddSpecializedMethodInHierarchy}
are composite refactoring operations described in~\cite{Ajouli-Cohen:2011}. \label{foot-operation}}

The resulting program is the one of
Fig.~\ref{base-prog:state3-visitor} which implements the
Visitor pattern.

   \subsection{Visitor to Composite}
      \label{FVisitToComp}

The base Visitor$\rightarrow$Composite transformation is
given in Fig.~\ref{fig-algo-retour}.
Again, the key point is to moving the business code back to
the composite hierarchy.
For this reason, we explain the whole algorithm again in
three stages: preparing the move, performing the move and
recovering the target structure.

        \paragraph{Steps I and II: Preparing for moving business code} 

%
Duplicate the method \xcode{accept(Visitor)}  in the whole composite
hierarchy into overloaded methods (see footnote~\ref{foot-operation}).
 Each overloaded method
takes one of the visitor classes as parameter
(step~\ref{visitor-composite-algo:specialize-accept}).
 These methods perform the same code as the
initial \xcode{accept(Visitor)} method, which is deleted since
its role is delegated to the new methods
(end of step~\ref{visitor-composite-algo:specialize-accept}).
At this point, all the invocations of the \emph{visit} method are done on subclasses of \xcode{Visitor}, so that we can
 delete the abstract
declaration of \emph{visit} methods from the abstract \xcode{Visitor}
(step~\ref{visitor-composite-algo:pushdown-visit}).
This will allow to inline visit methods in
composite classes (next step).

        \paragraph{Step III: Move business code to Composite classes}

Inline the \emph{visit} methods of the visitor classes and
delete them from these classes.  This boils down to moving
the business code back to composite classes (inside what was
delegators before). At this point, the visitor classes are
empty (no methods). They cannot be deleted now because they
are still referenced, they are deleted in the two last
steps.

        \paragraph{Steps IV to XI: Recovering Composite structure} 

In all the hierarchy, rename overloaded
methods \emph{accept} (from
step~\ref{visitor-composite-algo:specialize-accept}) to
methods with temporary different names to remove the
overloading: the
refactoring operation determines statically which instances
of the overloaded methods was referred to by the delegator,
and adjusts the invocation to the convenient deleguee
accordingly (step~\ref{visitor-composite-algo:rename-aux}).

           Then, remove the visitor parameters of these methods (step~\ref{visitor-composite-algo:remove-param}). Indeed, the parameter is not used, and the type of the parameter is not used anymore to resolve overloading.
 
Now we need to remove the delegation between the interface
for the business method (delegator) and the business code
(\emph{deleguee}, renamed at step~\ref{visitor-composite-algo:rename-aux}): 
 
  \begin{itemize}
   \item  Replace any recursive invocation of the deleguee methods by invocations of 
the corresponding method (step~\ref{visitor-composite-algo:fold}).
   \item  Push down the body of the delegators to the
sub-classes (step~\ref{visitor-composite-algo:pushdown-m}). This removes
the dynamic dispatch which would prevent future inlining.
   
    \item  To be able to delete the deleguee methods when they
will be inlined, delete their declarations from the abstract
super-class (step~\ref{visitor-composite-algo:pushdown-aux}).

     \item  Inline the deleguee methods in the concrete composites
classes and delete their declarations (step~\ref{visitor-composite-algo:inline-aux}).

      \item  Delete the visitor hierarchy since it is not used
anymore (steps~\ref{visitor-composite-algo:delete-visitors} and\ref{visitor-composite-algo:delete-visitor}).
  \end{itemize}

After performing this transformation, we find back the initial program of Fig.~\ref{base-prog:data} except a few changes in the 
layout and the comments.
%


\subsection{Precondition}
\label{sec:base-precondition}

These transformations are validated
in~\cite{Cohen-Ajouli-SAC} by inferring a minimum
precondition that ensures that
all the preconditions of successive atomic refactoring operations will
 be satisfied at run time, and that 
after performing a round-trip transformation, the resulting
 program is in a state which satisfies the initial
 precondition, so that the transformation can be applied
 again.

The computation of the minimum precondition is based on the
calculus of Kniesel and
Koch~\cite{composition-of-refactorings2004} and on a formal
description of the refactoring operations of \intellij. Those computed preconditions are valid under
the hypothesis that the formal description of the operations
are faithful with respect to the underlying tool (which has
been tested but not formally proven).

The full precondition (available in~\cite{Cohen-Ajouli-SAC}) is a conjunction of 49
predicates. 
Here is an extract of that precondition:

  \xcode{ExistsClass(Rectangle)}\\
$\wedge$ \xcode{ExistsMethodDefinition(Rectangle,show)}\\
$ \wedge \neg$\xcode{ExistsMethodDefinition(Rectangle,accept)}\\
$\wedge \neg$\xcode{ExistsMethodDefinition(Rectangle,printAux)}\\ $\wedge$ ...

For example, the proposition $\neg$\xcode{ExistsMethodDefinition(Rectangle, printAux)} is related to a temporary method name 
introduced in step~\ref{algo-composite-visitor:create-indirection} and indicates that such methods must not initially exist.  
%

%
%
In the following, we always consider the round-trip transformation for the computation of the preconditions.

\section{Pattern Variations}
\label{sect:variation}
The previous transformation algorithms can be applied only to
programs satisfying the computed preconditions.
In particular, the business methods must have no parameter,
must return \emph{void}, the superclass in the data type must
be an abstract class, and it must have only one level of
subclasses.
In the following, we relax these restrictions and show how
the transformations take these
changes into account.


%

%

\subsection{Methods with parameters}

\label{methods-WithParam}
  \subsubsection{Considered variation}
 We consider that  some business methods in the Composite structure have parameters, as exemplified by the following method \xcode{setColor} :

\begin{lstlisting}
// in Figure
abstract void setColor(int c)
\end{lstlisting}
\begin{lstlisting}
// in Rectangle
int color;
void setColor(int c) { this.color = c; }
\end{lstlisting}
\begin{lstlisting}
// in Group
void setColor(int c) {
   for (Figure child : children){child.setColor(c);} }
\end{lstlisting}

Note that the parameter \xcode{c} of the method \xcode{setColor} is passed to each recursive call (in the class \xcode{Group}).

\subsubsection{Target structure}

In the Visitor structure (Fig.~\ref{base-prog:state3-visitor}), the visitor object, which is
created by the interface methods of the class \xcode{Figure}, is passed recursively as parameter
of \emph{accept} and as receiver of \emph{visit}
invocations.
So, to take the parameter \xcode{c} into account, we put it into the state of that
visitor object, so that  it is available during the traversal:

 \begin{lstlisting}
class SetColorVisitor extends Visitor{

 final int c; 

 SetColorVisitor (int c){ this.c = c ; }  
  
 void visit(Rectangle r){ r.color = c; }
 
 void visit(Group g){ 
  for(Figure child : g.children){child.accept(this);} }}
\end{lstlisting}

The method \xcode{setColor} of the \xcode{Figure} abstract class  passes the parameter \xcode{c} to the constructor of the class \xcode{SetColorVisitor}, 
then passes the resulting visitor object (with \xcode{c} in its state) to the \emph{accept} method:

\begin{lstlisting}
// in Figure
void setColor(int c) { accept(new SetColorVisitor(c)); }
 \end{lstlisting}

The implementation of \emph{accept} in \xcode{Rectangle} and
\xcode{Group} is left unchanged.




  \subsubsection{Composite$\rightarrow$Visitor Transformation}

The refactoring operation of step~\ref{algo-composite-visitor:add-parameter} of the basic transformation (Fig.~\ref{fig-composite-visitor-transfo}) add a visitor parameter to the methods that becomes \emph{accept} later.
Here, we do not want to add the visitor parameter 
to the initial method parameter (such as \xcode{c}), 
but we want to replace the initial parameter with the visitor. To do that we apply the operation  \operation{IntroduceParameterObject} (step~\ref{algo-composite-visitor:add-parameter}.A below). Note that the refactoring operation 
\operation{IntroduceParameterObject} could not be used with methods without parameters.

For that reason, we distinguish methods with parameters and methods without parameters and we introduce the following notation to introduce different treatments in the transformation algorithm: 
\begin{itemize}
  \item $\mparam$: set of methods with parameters, here $\mparam =$ \{\xcode{setColor(int c)}\}.
  \item $\mwithoutparam$: set of methods without parameters, with $\mparam \cup \mwithoutparam = \mathbb{M}$ and  $\mparam \cap \mwithoutparam = \emptyset$.
\end{itemize}

Introducing a parameter object of type \xcode{A} to a method \mbox{\xcode{m(B b)}} for example creates a class A, moves the parameter b to A as an instance variable and finally 
changes  \xcode{m(B b)} into \xcode{m(A a)}. Any old access to \xcode{b} in the body of \xcode{m} will be replaced by \xcode{a.b}. 

The initial step~\ref{algo-composite-visitor:create-visitors}  
is  omitted for methods with parameters because the operation \operation{IntroduceParameterObject} creates the new class (step~\ref{algo-composite-visitor:create-visitors}.A below replaces step~\ref{algo-composite-visitor:create-visitors}). 

Here are the deviations from the basic algorithm for this variation:\\

\noindent\begin{boxedminipage}{\columnwidth}
\begin{enumerate}[4.A)]

\relsize{-1}
\sf

\item[\ref{algo-composite-visitor:create-visitors}.A)]  ForAll m in $\mwithoutparam$ do  
 \textbf{CreateEmptyClass}(\visitorclass{m})\\ \phantom{.} \hfill (replaces step \ref{algo-composite-visitor:create-visitors}) 
 
\item[\ref{algo-composite-visitor:add-parameter}.A)]   ForAll m in $\mparam$ do\\ \tab 
       \textbf{IntroduceParameterObject}(S, \auxname{m}, \visitorclass{m})\\[-2mm]

  ForAll m in $\mwithoutparam$ do\\ \tab 
       \textbf{AddParameterWithReuse}(S, \auxname{m}, \visitorclass{m})\\ \phantom{.}\hfill  (replaces step \ref{algo-composite-visitor:add-parameter})

\end{enumerate}

\end{boxedminipage}\\

\medskip

    \subsubsection{Visitor$\rightarrow$Composite Transformation}
Before deleting visitor classes (step~\ref{visitor-composite-algo:delete-visitors}) we have to check that there is no references to them in the Composite hierarchy. 
For the methods without parameters, we just remove the parameters corresponding to the visitor (step~\ref{visitor-composite-algo:remove-param}.A : restriction of  step ~\ref{visitor-composite-algo:remove-param} to methods without parameters) since at this moment those methods do not  
use that parameter.
 For example, at this moment (before step~\ref{visitor-composite-algo:remove-param}), the intermediate method for \xcode{print} in \xcode{Rectangle} is as follows:
\begin{lstlisting}
// in Rectangle
void printaux(PrintVisitor v){
       System.out.println("Rectangle");}
\end{lstlisting}

For the methods with parameters, instead of deleting the visitor parameter,  we have to  inline the occurrences of visitor classes to recover the initial parameter \xcode{c}. After applying step~\ref{visitor-composite-algo:inline-aux} (before deleting visitor classes), the method \xcode{setColor} is as follows:
\begin{lstlisting}
// in Rectangle
void setColor(int c){
       this.color = new SetColorVisitor(c).c;}
\end{lstlisting}

At this point we apply the operation \operation{InlineParameterObject} which will replace 
\xcode{\mbox{new SetColorVisitor(c).c}} by \xcode{c} (step~\ref{visitor-composite-algo:delete-visitors}.A), and then we can delete visitor classes (step~\ref{visitor-composite-algo:delete-visitor}). 

Here is the extension of the back transformation:\\[-2mm]

\noindent\begin{boxedminipage}{\columnwidth}
\begin{enumerate}[AA.A]

\relsize{-1}
\sf


\item[\ref{visitor-composite-algo:remove-param}.A]  
 ForAll m in $\mwithoutparam$ do \hfill (replaces step \ref{visitor-composite-algo:remove-param})\\ \tab \textbf{RemoveParameter}(S,\auxname{m},\visitorclass{m})

 \item[\ref{visitor-composite-algo:delete-visitors}.A]
 ForAll m in $\mparam$ do  \hfill (before step \ref{visitor-composite-algo:delete-visitors}) \\ \tab \textbf{InlineParameterObject}(S, \auxname{m}, \visitorclass{m})

\end{enumerate}
 
\end{boxedminipage}\\

\subsubsection{Computed Precondition}
%
The generated preconditions for this variations are related to the methods that have
parameters, such as
\xcode{ExistsMethodDefinitionWithParams(Figure,setColor,[int c])}. 
This constraint is imposed by the operation
\operation{IntroduceParameterObject}
(step~\ref{algo-composite-visitor:add-parameter}.A) since
that operation works only with methods with parameters (the set  $\mparam$ contains  \xcode{setColor(int c)}).

\subsection{Methods with different return types}
\label{methods-returning-result}
\subsubsection{Considered variation}
We consider now business methods with different return types. For example we consider a program 
with two methods: \xcode{Integer eval()} and \xcode{String  show()}.





\subsubsection{Target Structure}

Since we have methods with different return types,  we cannot use \xcode{void} to the \emph{accept} 
method. One solution is to have an \emph{accept} method variant for each return type by the means of overloading. 
But this breaks the beauty of the Visitor pattern (one \emph{accept} 
method for each business method instead of one \emph{accept} method to implement an abstract traversal).
 To avoid that, we use generic types as done in~\citet{Oliveira08}. 
In the abstract  class \xcode{Figure}, the \emph{accept} method becomes generic:
\begin{lstlisting}
abstract <T> T accept(Visitor <T> v)  
\end{lstlisting}

Note that the returned type is bound by the type of the
visitor class which appears as parameter. Each visitor class represents a business
method and its return type. The parameterized visitor structure is as follows:
\begin{lstlisting}
abstract class Visitor <T> {...}
\end{lstlisting}
\begin{lstlisting}
class EvalVisitor extends Visitor <Integer> {...} 
\end{lstlisting}
\begin{lstlisting}
class ShowVisitor extends Visitor <String> {...}
\end{lstlisting}



\paragraph{Remark} Because of the restriction in the use of generic types in
Java, returned types which are raw types, such as \xcode{int}
or \xcode{bool}, must be converted to object types
such as \xcode{Integer} or \xcode{Boolean}.
In the case of \xcode{void}, one can use \xcode{Object} and add a \xcode{return
null} statement (we use a refactoring operation to do that).

\subsubsection{Composite$\rightarrow$Visitor Transformation}
We use the following notations in the algorithm corresponding to this variation:

\begin{itemize}
   \item $\mathbb{R}$: Set of methods and their corresponding return types, here $\mathbb{R} = $\{(\xcode{show},\xcode{String}), (\xcode{eval},\xcode{Integer})\}.
\end{itemize}
In step~\ref{algo-composite-visitor:superclass} of the basic algorithm, the operation \operation{ExtractSuperClass} creates a new abstract class and 
pulls up abstract declarations of visit methods. In the considered variation, we have to use an extension of the pull up operation that introduces generic types 
in the super class to be able to insert abstract declarations for methods with different return types.

To deal with this variation we apply the operation \operation{ExtractSuperClassWithoutPullUp} then the operation \operation{PullUpWithGenerics}\footnote{\url{http://plugins.jetbrains.com/plugin/?idea\_ce&id=6889}}
instead 
of the operation \operation{ExtractSuperClass} of the step~\ref{algo-composite-visitor:superclass} (step~\ref{algo-composite-visitor:superclass}.B).\\

\noindent\begin{boxedminipage}{\columnwidth}

\begin{enumerate}
\relsize{-1}
\sf

\item[\ref{algo-composite-visitor:superclass}.B]
 \textbf{ExtractSuperClassWithoutPullUp}($\mathbb{V}$, "Visitor") ;\\
          ForAll m in $\mathbb{M}$, c in $\mathbb{C}$ do \\ \tab 
            \textbf{PullUpWithGenerics}(\visitorclass{m}, "visit","Visitor")  \hfill (replaces~\ref{algo-composite-visitor:superclass}) 


\end{enumerate}

\end{boxedminipage}\\

\subsubsection{Visitor$\rightarrow$Composite Transformation}

At the step~\ref{visitor-composite-algo:specialize-accept} of the base algorithm, 
we must specify the return type 
of each \emph{accept} method. The convenient return types could be identified directly from return types of visit methods existing in concrete visitors. 
This is done by the operation \operation{AddSpecialisedMethodWithGenerics} (step~\ref{visitor-composite-algo:specialize-accept}.B).\\

\noindent\begin{boxedminipage}{\columnwidth}

\begin{enumerate}
\relsize{-1}
\sf

\item[\ref{visitor-composite-algo:specialize-accept}.B] ForAll v in $\mathbb{V}$ do\\ \tab
 \textbf{AddSpecializedMethodWithGenerics}(S,"accept",$\mathbb{R}$,\\"Visitor",v) 
 \hfill (replaces~\ref{visitor-composite-algo:specialize-accept}) 

  
\end{enumerate}

\end{boxedminipage}\\

\subsubsection{Computed Precondition}

The only difference with the basic preconditions is the check of return types which should not be raw types. This precondition is required by 
the operation \operation{PullUpWithGenerics}.


\subsection{Hierarchy With multilevel}
  \label{multilevel-variant}
\subsubsection{Considered variation}
   We consider that the Composite hierarchy has multiple
levels, with a random repartition of business code: some
business methods are inherited, and some other are
overridden.

For example, we consider the class \xcode{Rectangle} has a subclass \xcode{ColoredRectangle} where the method \xcode{print} is  overridden whereas 
the second method \xcode{show} is inherited:
\begin{lstlisting}
class ColoredRectangle extends Rectangle{ 
 int color; 
 ColoredRectangle (int c){ this.color = c; } 

 void print{System.out.println(
       "Rectangle colored with " + color); } }
\end{lstlisting}

\subsubsection{Target Structure}

In order to have  in visitor classes one \emph{visit} method for each class
of the Composite hierarchy,
the code of the method \xcode{show()} defined in
\xcode{Rectangle} in the Composite structure and inherited
by \xcode{ColoredRectangle}, is placed in the methods
\xcode{\mbox{visit(ColoredRectangle c)}} and
\xcode{\mbox{visit(Rectangle r)}} in \xcode{ShowVisitor}:
%
%

%
%

\begin{lstlisting}
class ShowVisitor extends Visitor{
 void visit(Group g){...}
 
 void visit(Rectangle r){
            System.out.println("Rectangle :"+ r +".");}

 void visit(ColoredRectangle c){
           System.out.println("Rectangle :"+ c +".");}}
  
\end{lstlisting}

\subsubsection{Composite$\rightarrow$Visitor Transformation}

In order to push down a duplicate of the inherited method to the right 
subclass, we apply the operation \operation{PushDownCopy}
(step~\ref{algo-composite-visitor:create-visitors}.C) before running the basic algorithm.

We use the following notations in the algorithm corresponding to this variation:

\newcommand{\getinherited}[1]{\ensuremath{i(\mbox{#1})}}
\newcommand{\getsuper}[1]{\ensuremath{s(\mbox{#1})}}

\begin{itemize}
\item $\getinherited{$c$}$: a function that gives the list of inherited methods of a class ; 
here $\getinherited{\xcode{ColoredRectangle}}$ = \{\xcode{show()}\}.
 \item  $\getsuper{$c$}$: a function that gives the superclass of a class.\\

\end{itemize}

\noindent\begin{boxedminipage}{\columnwidth}
\begin{enumerate}
 \relsize{-1}
\sf

\item[\ref{algo-composite-visitor:create-visitors}.C] ForAll c in $\mathbb{C}$,
  ForAll m in $\getinherited{$c$}$ do  \hfill (before~\ref{algo-composite-visitor:create-visitors})\\ \tab \tab
   \textbf{PushDownCopy}(c,m,$\getsuper{$c$}$) 

 
\end{enumerate}

\end{boxedminipage}\\

\subsubsection{Visitor$\rightarrow$Composite Transformation}
 First we apply the basic algorithm. Then, in order to get back the initial structure
 we delete methods (step~\ref{visitor-composite-algo:delete-visitor}.C) that were 
initially added in these classes in the step~\ref{algo-composite-visitor:create-visitors}.C of the forward transformation.\\

\noindent\begin{boxedminipage}{\columnwidth}
\begin{enumerate}[XX.X]

\relsize{-1}
\sf

\item[\ref{visitor-composite-algo:delete-visitor}.C]  ForAll (c,m) in $\mathbb{C}$,
  ForAll m in $\getinherited{$c$}$  do \\ \tab \tab 
    \textbf{DeleteMethod}(c,m)   \hfill (after~\ref{visitor-composite-algo:delete-visitor})
\end{enumerate}

\end{boxedminipage}\\


\paragraph{Remark} 
The refactoring operation that performs that deletion should
rely on the fact that the code of the deleted method is the
same as the code in the super class.
%

\subsubsection{Computed Precondition} 
 The precondition that characterizes the transformation for this variation is: the method \xcode{show}  must not be overridden in  the class \xcode{ColoredRectangle} and must not call any 
overloaded method called with the argument \xcode{this} in the class  \xcode{Rectangle}. This is due to the fact that, if the method \xcode{Rectangle::show} have any overloaded method 
with the argument \xcode{this}, we could not copy this method to the class \xcode{ColoredRectangle} since the argument \xcode{this} will refer to that class and which could change the behavior of this 
method which is supposed to keep the same behavior either in the class \xcode{Rectangle} or in the class \xcode{ColoredRectangle}.



%

\subsection{Interface instead of Abstract Class}
   \label{interface-variant}

\subsubsection{Considered Variation}
 We now consider that the root of the Composite hierarchy is
 not an abstract class but an interface and that
there is 
an intermediary abstract class between it and its subclasses. 
This architecture is found in real softwares: libraries are often provided by the means of an interface and compiled byte-code (\emph{Facade} pattern). 

 We suppose that there are no other subclasses 
implementing the interface. 
%
%


\begin{lstlisting}
interface Figure{
 void print();
}
\end{lstlisting}
\begin{lstlisting}
abstract class AbstractFigure implements Figure {
 abstract void print();
}
\end{lstlisting}
\begin{lstlisting}
class Group extends AbstractFigure{
 
ArrayList<Figure> children = ...

 void print(){
     ...
     for(Figure child : children){ child.print();}}
}

\end{lstlisting}
\subsubsection{Target Structure}


Here is a possible target structure corresponding to the considered variation:
\begin{lstlisting}
interface Figure{
 void print();
 void accept(Visitor v);
}
\end{lstlisting}
 \begin{lstlisting}
abstract class AbstractFigure implements Figure{
 void print(){accept(new PrintVisitor());}
}
\end{lstlisting}
\begin{lstlisting}
class Group extends AbstractFigure{ 
 ArrayList<Figure> children = ...
 void accept(Visitor v){v.visit(this);}
}
\end{lstlisting}
 \begin{lstlisting}
class PrintVisitor extends Visitor{
  void visit(Group g){
     for(Figure child : g.children){ child.accept(this);}}
}
\end{lstlisting}

Note that the loop in \xcode{visit(Group)} is done on objects of type \xcode{Figure} (not \xcode{AbstractFigure}).

\subsubsection{Composite$\rightarrow$Visitor Transformation}

To reach the target structure, we have to create a delegator
\xcode{print()\{printaux(..)\}} in the class
\xcode{AbstractFigure}
 and
inline the recursive call of \xcode{print} in \xcode{Group}
(steps~\ref{algo-composite-visitor:create-indirection} and~\ref{algo-composite-visitor:inline-methods}).
But that recursive call refers to the method
\xcode{print} declared in the \xcode{Figure} interface whereas
the delegator is defined in the abstract class \xcode{AbstractFigure}.
To solve that, we introduce a downcast to the class
\xcode{AbstractFigure} in the recursive call to
\xcode{print} as follows: \xcode{((AbstractFigure)
  child).print()}
(step~\ref{algo-composite-visitor:inline-methods}.D).
This makes the inlining by the refactoring tool possible.
%
%
%
This downcast is legal because we suppose that the interface
has no other implementation than the abstract class.

%
After creating the method \xcode{accept} (step~\ref{algo-composite-visitor:merge}),
we pull up its declaration to the
interface \xcode{Figure}, then we delete
the downcast (step~\ref{algo-composite-visitor:merge}.D).

\noindent\begin{boxedminipage}{\columnwidth}
\begin{enumerate}

\relsize{-1}
\sf

\item[\ref{algo-composite-visitor:inline-methods}.D] ForAll m in $\mathbb{M}$, c in $\mathbb{C}$ do\\ \tab \tab
  \textbf{IntroduceDownCast}(c,m,S)  \hfill (before~\ref{algo-composite-visitor:inline-methods})

\item[\ref{algo-composite-visitor:merge}.D] 
 \textbf{pullupAbstractMethod}(S, "accept", I) 

ForAll v in $\mathbb{V}$ do\\ \tab \tab
  \textbf{DeleteDownCast}(v,"accept")  \hfill (after~\ref{algo-composite-visitor:merge})

\end{enumerate}

\end{boxedminipage}\\

\paragraph{Real practice of the transformation}
The algorithms shown above represent the ideal solution to get a Visitor structure.
In fact, there is no operation in the refactoring tools
we use to manage downcasts.
In order to automate the full transformation, we do not use
downcasts and do not inline the delegator.
As a result we get a Visitor with indirect  recursion as follows:

\begin{lstlisting}
// In Figure
   void print();
\end{lstlisting}
\begin{lstlisting}
// In AbstractFigure
   abstract void accept(Visitor v);
   void print(){accept(new PrintVisitor();}
\end{lstlisting}
\begin{lstlisting}
// In PrintVisitor
   void visit(Group g){
     for(Figure child : g.children){ child.print();}}
\end{lstlisting}

We can see that at each recursive invocation a new instance
of a Visitor is created. The result is legal but shows a poor
use of memory. 
This problem disappears when the initial
Composite structure is recovered. 
Moreover, if needed, the downcast can be introduced manually
(or the refactoring operation can be implemented).


So, in practice, the variation in the algorithm is: do not apply step \ref{algo-composite-visitor:inline-methods} (nor \ref{algo-composite-visitor:inline-methods}.D); do not apply step \ref{algo-composite-visitor:merge}.D (but  step \ref{algo-composite-visitor:merge}).

\subsubsection {Visitor$\rightarrow$Composite Transformation}

After the \emph{practical} Composite$\rightarrow$Visitor
transformation, the base Visitor$\rightarrow$Composite
transformation can be applied without performing the step~\ref{visitor-composite-algo:fold}.

After the full Composite$\rightarrow$Visitor transformation
described above (with downcasts), we also have to add and
remove some downcasts to recover the Composite structure (before step~\ref{visitor-composite-algo:fold} and after~\ref{visitor-composite-algo:fold}, \emph{the detail is not given for reason of space}).

\subsubsection{Computed Precondition} 
We consider the practical algorithms without downcasts. 
Because we do not apply inlinings in
step~\ref{algo-composite-visitor:inline-methods}, the
constraints \xcode{IsRecursiveMethod(Group,print)} and
  \xcode{IsRecursiveMethod(Group,show)} have disappeared
    from the computed minimum precondition.

\subsection{Support for Precondition Generation}

To generate the minimum preconditions to ensure the
correctness of our transformations, we described 24
refactoring operations with 480 \emph{backward description}
rules (we use the concept of \emph{backward descriptions}
from the work of Kniesel and  Koch~\cite{composition-of-refactorings2004}). 
The specification of each 
refactoring operation (preconditions and \emph{backward descriptions}) are given in~\cite{Ajouli-Cohen:2011}.


\section{Use Case : JHotDraw}
\label{usecase}
In this section we we apply our transformation to the JHotDraw framework.

   \subsubsection{Overview}
%

In JHotDraw, there is a Composite structure with 
18 classes and 6 business methods which shows the four variations presented above. We aliment the transformation algorithm with the following data:

\begin{itemize}

\item  $S = $ \xcode{AbstractFigure}.
\item $\mathbb{C} = $ \{  \xcode{EllipseFigure}, \xcode{DiamondFigure}, \xcode{RectangleFigure},  \xcode{RoundRectangleFigure},  \xcode{TriangleFigure}, \xcode{TextFigure},  \xcode{BezierFigure}, 
 \xcode{TextAreaFigure}, ...
\}.

\item $\mparam = $ \{ \xcode{basicTransform (AffineTransform tx)}, \xcode{contains(Point2D.Double p)},  \xcode{setAttribute(AttributeKey key,Object value)}, \xcode{findFigureInside(Point2D.Double p)},
 \xcode{addNotify(Const "Drawing d)},\xcode{removeNotify(Drawing d)}\}.

\item $\mwithoutparam = \emptyset$.

\item $\mathbb{R} = $ \{ (\xcode{basicTransform},\xcode{Void}), (\xcode{contains},\xcode{Boolean}),  (\xcode{setAttribute}, \xcode{Void}), (\xcode{findFigureInside},\xcode{Figure}),
 (\xcode{addNotify}, \xcode{Void}), (\xcode{removeNotify}, \xcode{Void})\}.

\item $\getsuper{\xcode{LineConnectionFigure}} $ =  \{\xcode{BezierFigure}\}\\ $\getsuper{...} $=  ...  

\item $\getinherited{\xcode{LineConnectionFigure}}$ =  \{\xcode{findFigureInside}, 
                                    \xcode{setAttribute},\xcode{contains}\},\\ $\getinherited{...}$ = ...





\end{itemize}

%



%

 \subsubsection{From Composite to Visitor}
  
To switch from the Composite structure of JHotDraw to its Visitor structure we  apply the following sequence of steps: \ref{algo-composite-visitor:create-visitors}.C ; 
\ref{algo-composite-visitor:create-indirection} ; \ref{algo-composite-visitor:add-parameter}.A ; \ref{algo-composite-visitor:move} ; \ref{algo-composite-visitor:superclass}.B ; \ref{algo-composite-visitor:generalise-parameter} ; \ref{algo-composite-visitor:merge}.

 \subsubsection{From Visitor to Composite} To recover the initial structure, we apply the following: steps~\ref{visitor-composite-algo:specialize-accept}.B ; 
\ref{visitor-composite-algo:delete-method-in-hierarchy};
\ref{visitor-composite-algo:pushdown-visit};  \ref{visitor-composite-algo:inline-visit} ; \ref{visitor-composite-algo:rename-aux} ; \ref{visitor-composite-algo:remove-param}.A; 
\ref{visitor-composite-algo:pushdown-m} ; \ref{visitor-composite-algo:pushdown-aux} ; \ref{visitor-composite-algo:inline-aux} ; 
 \ref{visitor-composite-algo:delete-visitors}.A ; \ref{visitor-composite-algo:delete-visitors} ; \ref{visitor-composite-algo:delete-visitor} ; \ref{visitor-composite-algo:delete-visitor}.C.

 
\subsubsection{Generated Precondition}     
We have computed a minimum precondition that ensures the correctness of the round-trip transformation. 
That precondition, given in~\cite{Ajouli-Cohen:2011}, is a conjunction of 1852 propositions.

\section{Related work}
 
\paragraph{Transforming Design Pattern Implementations}

\citet{Smalltalk-refactoring-1997} use
sequences of basic refactoring operations to introduce
design patterns in existing programs, including the Visitor pattern. 
%
%
\citet{Cinneide00compositerefactorings} provide automatic preconditions
generation for such sequences, without considering Visitors however. 
%
%
%
%
%
\citet{Hills:2011, Hills:2012} implement the reverse transformation:
they remove the Visitor structure from a real interpreter.
 Kerievsky~\cite{Kerievsky04} provides a catalog of
 guidelines to introduce design patterns by refactoring
 sequences, including the Visitor.
%

%
%
%

\paragraph{Variations in Composite and Visitor implementations}


%
\citet{Kerievsky04} introduces Visitors in two
variations of class hierarchies.
The first one is close to the base architecture we consider
(Sec.~\ref{sect:basictrasnformation}), but misses some
features of the Composite pattern: 1) the business methods
initially do not have abstract declarations in the superclass
of the hierarchy and 2) there is no recursion in the class
structure.
Moreover, only one business method is considered (3).
As a result, there are three differences between Kerievsky's
algorithm and our base algorithm: \begin{enumerate} 

\item Our steps~\ref{algo-composite-visitor:create-indirection} and~\ref{algo-composite-visitor:merge} that make the \emph{accept} method appear are done in a different way in 
\citet{Kerievsky04}.

\item 
Without recursion, our step~\ref{algo-composite-visitor:inline-methods} is
pointless, and in
step~\ref{algo-composite-visitor:add-parameter}  the ``\emph{any var}'' option of
the \textbf{AddParameter} operation is also useless (see \textbf{AddParameterWithReuse}
in~\cite{Ajouli-Cohen:2011}).

\item  Only one
Visitor class is introduced in Kerievsky, so that he does not need to add
an interface for the Visitor hierarchy (steps 6 to 8 in our
algorithm).

\end{enumerate}

Also, the algorithms of Kerievsky are only generic guidelines
and are not formally analysed.

The second variation takes
place in a very different program, so that comparing the
algorithms is not relevant.

\section{Conclusion}


%

The contributions of this article are the following:

\begin{itemize}
 
\item We have selected four common variations in the
   implementation of the Composite pattern and we have shown
   how these variations reflect in the 
   Visitor pattern.

\item
For each variation, we have extended the
previously defined transformation. The resulting transformations are automated and invertible.

\item For each variation, we have checked the validity of
  the adapted transformations by computing the minimum
  precondition that ensures their success.
%

 \item We have applied the four extensions of the algorithms
   on a real-size use case. The resulting algorithm is also
   validated by computing its precondition.

\item
This use case also shows that the algorithm extensions can be used
together without conflicting interaction between them.

\end{itemize}

The potential impact of this work in  software industry
could be estimated with the following future work:
\begin{itemize}

\item Estimate the repartition of variations of
  Composite and Visitor patterns in industrial softwares and
  compare it to the variations we have considered. 
See which uncovered variations are important in industry.

 \item Estimate the value of changing the structure of
   a source code in order to ease its
   maintenance by experimenting it with real programmers, on real
   programs.

\end{itemize}

Of course, to be applicable in industry, we also have to
provide some refactoring operations that are currently
missing (see sections~\ref{multilevel-variant} and~\ref{interface-variant}).

Besides the potential industrial benefits, studying pattern
variations and their impact in dual architectures has a
pedagogical interest: it can be used to better understand the
patterns themselves~\cite{Kerievsky04}.
In particular, the Visitor pattern is not trivial to
understand. 
This work
shows for instance how visitors with or without a state are
related to methods with or without parameters (variation A) and
how generic types are needed in visitors
 when the business methods have different return types (variation B).
%
%



As a future work, it would be usefull to use refactoring
inference tools such as~\cite{Moghadam-Cinneide:2012} to
find changes in the transformations automatically.

\newpage
\bibliographystyle{abbrvnat}



\end{document}